\documentclass[prd,aps,showpacs,floats,letterpaper,floatfix,groupedaddress,eqsecnum,nofootinbib]{revtex4}

\usepackage{graphicx,epsf, epsfig, amssymb}
\usepackage{bm}
\def\be{\begin{equation}}
\def\ee{\end{equation}}
\def\beq{\begin{eqnarray}}
\def\eeq{\end{eqnarray}}
\begin{document}

\title{Hawking emission of gravitons in higher dimensions: non-rotating black holes}

\author{Vitor Cardoso} \email{vcardoso@phy.olemiss.edu}
\affiliation{Dept. of Physics and Astronomy, The University of
Mississippi, University, MS 38677-1848, USA \footnote{Also at Centro
de F\'{\i}sica Computacional, Universidade de Coimbra, P-3004-516
Coimbra, Portugal}}

\author{Marco Cavagli\`a} \email{cavaglia@phy.olemiss.edu}
\affiliation{Dept. of Physics and Astronomy, The University of
Mississippi, University, MS 38677-1848, USA}

\author{Leonardo Gualtieri} \email{gualtieri@roma1.infn.it}
\affiliation{Centro Studi e Ricerche E. Fermi, Compendio Viminale, 00184 Rome, Italy and\\
Dipartimento di Fisica Universit\`a di Roma ``La Sapienza''/Sezione INFN Roma1,
P.le A. Moro 2, 00185 Rome, Italy}

\date{\today}


\begin{abstract}

We compute the absorption cross section and the total power carried by gravitons in the evaporation process of
a higher-dimensional non-rotating black hole. These results are applied to a model of extra dimensions with
standard model fields propagating on a brane. The emission of gravitons in the bulk is highly enhanced as the
spacetime dimensionality increases. The implications for the detection of black holes in particle colliders and
ultrahigh-energy cosmic ray air showers are briefly discussed.

\end{abstract}

\maketitle

\section{Introduction}
Models with extra dimensions have emerged as the most successful candidates for a consistent
unified theory of fundamental forces. The most recent formulation of superstring theory suggests
that the five consistent string theories and 11-dimensional supergravity constitute special points
in the moduli space of a more fundamental nonperturbative theory, called $M$-theory
\cite{Polchinski:1998rq}. In models of warped or large extra dimensions, standard model fields
but the graviton are confined to a three-dimensional membrane \cite{Arkani-Hamed:1998rs,
Randall:1999vf}. The observed hierarchy between the electroweak and the gravitational coupling
constants is naturally explained by the largeness of the extra dimensions.

The presence of extra dimensions changes drastically our understanding of high-energy physics and
gravitational physics. Gravity in higher dimensions is much different than in four dimensions. For
example, black holes with a fixed mass may have arbitrarily large angular momentum \cite{Myers:1986un}.
The uniqueness theorem does not hold, allowing higher-dimensional black holes with non-spherical
topology. A n\"{a}ive analysis suggests that higher-dimensional black holes should evaporate
\cite{Hawking:1974sw} more quickly than in four dimensions, due to the larger phase space. Moreover,
brane emission should dominate over bulk emission because standard model fields carry a larger number
of d.o.f.\ than the graviton \cite{Emparan:2000rs}. However, a black hole does not radiate exactly as a
black body and the emission spectrum depends crucially on the structure and dimensionality of the
spacetime. A large graviton emissivity could reverse the above conclusion; if the probability of
emitting spin-2 quanta is much higher than the probability of emitting lower spin quanta, the black
hole could evaporate mainly in the bulk \cite{Cavaglia:2003hg}. A conclusive statement on this issue
can only be reached if the relative emissivities of all fields (greybody factors) are known. This is
particularly relevant for the phenomenology of black holes in models of low-energy scale gravity
\cite{Cavaglia:2002si,Kanti:2004nr}. Unequivocal detection of subatomic black holes in particle
colliders \cite{Banks:1999gd} and ultrahigh-energy cosmic ray observatories \cite{Feng:2001ib,
Ahn:2005bi} is only possible if a consistent fraction of the initial black hole mass
\cite{Eardley:2002re,Cardoso:2002ay} is channeled into brane fields.

If the center-of-mass energy of the event is sufficiently large compared to the Planck scale, quantum
gravitational effects can be neglected and the black hole can be treated classically. It is commonly
accepted that black holes with masses larger than few Planck masses satisfy this criterion. Under this
assumption, if the fundamental gravitational scale is about a TeV, micro black holes produced at the
Large Hadron Collider (center-of-mass energy = 14 TeV) and in ultrahigh-energy cosmic ray showers
(center-of-mass energy $\gtrsim 50$ TeV) can be considered classical. Throughout the paper we will
assume that this is the case. (See Ref.~\cite{Ahn:2005bi} and references therein for a discussion of
the uncertainties deriving from this assumption.)

The relative emissivities per degree of freedom (d.o.f.) of a classical four-dimensional non-rotating black
hole are 1, 0.37, 0.11 and 0.01 for spin-0, -1/2, -1 and -2, respectively \cite{Page:1976df}. In that case, the
graviton power loss is negligible compared to the loss in other standard model channels. Since brane fields are
constrained in four dimensions, the relative greybody factors for these fields approximately retain the above
values in higher dimensions. The emission rates for the fields on the brane have been computed in Ref.\
\cite{Kanti:2002nr}. The graviton emission is expected to be larger in higher dimensions due to the increase in
the number of its helicity states. The authors have recently calculated the exact absorption cross section,
power and emission rate for gravitons in generic $D$-dimensions \cite{Cardoso:2005vb}. (See also Ref.\
\cite{Cornell:2005ux}.) The purpose of this paper is to discuss these results in more detail. The graviton
emissivity is found to be highly enhanced as the spacetime dimensionality increases. Although this increase is
not sufficient to lead to a domination of bulk emission over brane emission, at least for the standard model, a
consistent fraction of the higher-dimensional black hole mass is lost in the bulk.

The organization of this paper is as follows. In Section \ref{sec:eqs} we fix notations and
briefly review the basics of gravitational perturbations in the higher-dimensional black hole
geometry. In Section \ref{sec:abscs} we derive the field absorption cross sections from the
absorption probabilities. Details of this derivation are included in the Appendix. The low-energy
absorption probabilities and the cross sections for spin-0, -1 and -2 fields in generic dimensions
are computed in Section \ref{sec:lowen}. This derivation is valid for scalars, vectors and
gravitons in the bulk and generalizes previous results \cite{Kanti:2002nr}. In Section
\ref{sec:highen} we prove that the high-energy behavior of the cross section is universal and
reduces to the capture cross section for a point particle. Numerical results for the total power
and the emission rates for all known particle species are obtained in Section \ref{sec:final}.
Finally, Section \ref{sec:concl} contains our conclusions.
\section{\label{sec:eqs}Equations and conventions}
The formalism to handle gravitational perturbations of a higher-dimensional non-rotating black hole
was developed by Kodama and Ishibashi \cite{Kodama:2003jz} (hereafter, KI). In this section we
briefly review their main results.
\subsection{\label{subsec:hel}Number of degrees of freedom of gravitational waves}
In four dimensions, gravitational waves have two possible helicities \cite{mtw}, corresponding to
the number of spatial directions transverse to the propagation axis. In generic $D$ dimensions,
gravitational waves can be described by a symmetric traceless tensor of rank $D-2$, corresponding
to the $\Box\!\Box$ representation of $SO(D-2)$. The number of helicities is
\begin{equation}
{\cal N}=\frac{(D-2)(D-1)}{2}-1=\frac{D(D-3)}{2}\,.
\label{defN}
\end{equation}
This representation is decomposed into tensor ($T$), vector ($V$), and scalar ($S$) perturbations
on the sphere $S^{D-2}$. These components correspond to symmetric traceless tensors, vectors and
scalars, respectively. Tensors and vectors are divercenceless on $S^{D-2}$. The number of d.o.f.\
are
\be
{\cal N}_T=\left(\frac{(D-2)(D-1)}{2}-1\right)-(D-2)\,,\qquad
{\cal N}_V=\left(D-2\right)-1\,,\qquad
{\cal N}_S=1\,.
\ee
The perturbations are further expanded in tensor, vector and spherical harmonics on $S^{D-2}$. The
total number of helicity states is obtained by considering the multiplicities of these components.
A massless particle of spin $J$ in four dimensions has helicities $+J$ and $-J$, corresponding to
the projections of the spin along the direction of motion. These states provide a representation of
the little group $SO(D-2)=SO(2)$, i.e.\ the group of spatial rotations preserving the particle
direction of motion. In four dimensions all massless particles have two helicities since all
non-singlet representations of $SO(2)$ have dimension two. The description of spin in $D$
dimensions proceeds similarly to the four-dimensional case. For instance, in five dimensions there
are three directions orthogonal to the direction of motion. The little group is $SO(3)$ and the
helicities of the 5-dimensional graviton are $2\,,1\,,0\,,-1\,,-2$. For a general discussion, see
Ref.\ \cite{Slansky}.
\subsection{\label{sec:metric} Metric and master equations for gravitational perturbations}
The metric of the higher-dimensional non-rotating spherically symmetric black hole is
\cite{Tangherlini,Myers:1986un}
\begin{equation}
ds^2= -fdt^2+ f^{-1}dr^2 +r^2d\Omega_{D-2}^2\,,
\label{metrictang}
\end{equation}
where
\begin{equation}
f=1-\frac{r_H}{r^{D-3}}\,.
\label{fdef}
\end{equation}
The mass of the black hole is $M=(D-2)\Omega_{D-2}\,r_H/(16\pi{\cal G})$, where $\Omega_{D-2}$ is
the volume of the unit $(D-2)$-dimensional sphere with line element $d\Omega_{D-2}^{2}$ and
${\cal G}$ is Newton's constant in $D$ dimensions. Withouth loss of generality, we choose
$r_H=1$. This amounts to rescaling the radial coordinate and $\omega$. Hereafter, $\omega$ stands
for $\omega\,r_H$.

Let us consider weak perturbations of the geometry (\ref{metrictang}) by external fields. If the
field is a scalar, the contribution to the total energy-momentum tensor is negligible. The
evolution for the field is simply given by the Klein-Gordon equation in the fixed background. The
evolution equations for the electromagnetic field were derived by Crispino, Higuchi and Matsas
\cite{Crispino:2000jx} and more recently by KI in the context of charged black hole
perturbations. According to the KI formalism, the perturbations are divided into vector and
scalar perturbations. The gravitational evolution equations were derived by KI
\cite{Kodama:2003jz} following earlier work by Regge and Wheeler \cite{Regge} and Zerilli
\cite{Zerilli:1970se} in four dimensions. The gravitational perturbations are divided in scalar,
vector and tensor perturbations. Tensor perturbations exist only in $D>4$.

The evolution equation for all known fields (scalar, electromagnetic and gravitational) can be
reduced to the second order differential equation
\begin{equation}
\frac{d^2\Psi}{dr_{*}^2}+(\omega^2-V)\Psi=0 \,,
\label{eveq}
\end{equation}
where $r$ is a function of the tortoise coordinate $r_*$, which is defined by $\partial r/\partial
r_*=f(r)$. With the exception of gravitational scalar perturbations, the potential $V$ in Eq.\
(\ref{eveq}) can be written as
\be V = f(r) {\biggl [} \frac{l(l+D-3)}{r^2}+\frac{(D-2)(D-4)}{4r^2}+
\frac{(1-p^2)(D-2)^2}{4r^{D-1}} {\biggr ]}\,.
\label{potentialj}
\ee
The constant $p$ depends on the field under consideration:
\begin{equation}
p=\left\{ \begin{array}{llll} 0\ & \hbox{for scalar and gravitational tensor perturbations,}\\
2 & \hbox{for gravitational vector perturbations,}\\
2/(D-2) & \hbox{for electromagnetic vector perturbations,}\\
2(D-3)/(D-2) &\hbox{for electromagnetic scalar perturbations.}
\end{array}\right.
\label{jdef}
\end{equation}
For radiative modes, the angular quantum number $l$ takes the integer values
\begin{equation}
l=\left\{ \begin{array}{ll}
0\,,1\dots & \hbox{for scalar perturbations,}\\
1\,,2\dots & \hbox{for electromagnetic perturbations,}\\
2\,,3\dots & \hbox{for gravitational perturbations.}
\end{array}\right.
\label{ldef}
\end{equation}
The potential for gravitational scalar perturbations is
\be V= f\frac{Q(r)}{16r^2H(r)^2}\,,
\label{def-potential}
\ee
where
\begin{eqnarray}
Q(r)&=&(D-2)^4(D-1)^2x^3+(D-2)(D-1) \\ \nonumber &\times&\{4[2(D-2)^2-3(D-2)+4]m+(D-2)(D-4)(D-6)(D-1)\}x^2 \\
\nonumber &-&12(D-2)\{(D-6)m+(D-2)(D-1)(D-4)\}m x +16m^3+4D(D-2)m^2\,, \\ H(r)&=&m+{1\over
2}(D-2)(D-1)x\,,\quad m=l(l+D-3)-(D-2)\,,\quad x \equiv \frac{1}{r^{D-3}}\,,
\end{eqnarray}
and $l\ge 2$ has been assumed. In four dimensions, the equations for scalar and vector
perturbations reduce to the Zerilli and Regge-Wheeler equations, respectively.
In the limit $l\rightarrow \infty$, Eq.\  (\ref{eveq}) shows the universal behavior
\be \frac{d^2\Psi}{dr_{*}^2}+(\omega^2-f\frac{l^2}{r^2})\Psi=0
\,,\quad l\rightarrow \infty \,.
\ee
The above equation guarantees that all perturbations behave identically in this limit. The
effective potential for tensor perturbations is equal to the potential of a massless scalar
field in the higher-dimensional Schwarzschild black hole background \cite{Cardoso:2002pa}.

Assuming a harmonic wave $e^{i\omega t}$ and ingoing waves near the horizon, we
impose the boundary condition
\be \Psi(r)\rightarrow e^{- i\omega r_*} \sim (r-1)^{-
i\omega/(D-3)}\,,\quad r \rightarrow r_+\,.
\label{def-bcrmais}
\ee
At the asymptotic infinity, the wave behavior includes both ingoing and outgoing waves
\be
\Psi(r)\rightarrow T e^{-i\omega r_*}+ R e^{i\omega r_*}\,,\quad
r_*\rightarrow \infty\,.
\label{def-bc}
\ee
The physical picture is a wave with amplitude $T$ scattering on the black hole. Part of this wave
is reflected back with amplitude $|R|^2$  and part is absorbed with probability $|{\cal
A}|^2=(|T|^2-|R|^2)/|T|^2$.
\section{\label{sec:abscs}From absorption probabilities to absorption cross sections}
The rate of absorbed particles for a plane wave of spin $s$ and flux $\Phi^s$ is
\be
\frac{dN_{s}}{dt}=\sigma^{s}\Phi^{s}\,,\label{defsigma}
\ee
where $\sigma^s$ is the absorption cross section. In Eq.\ (\ref{defsigma}) we sum over all final
states and average over the initial states. The total cross section is obtained by summing the
absorption coefficients for each single mode $l$, ${\cal A}^{s}_{l}$, weighted by the multiplicity
factors. For a scalar field, the result is \footnote{See, for example, Ref.\ \cite{Unruh:1976fm}
for the four-dimensional case and the Appendix \ref{crossapp} and Ref.\
\cite{Gubser:1997qr,Kanti:2004nr} for its $D$-dimensional generalization.}
\be
\sigma^{s=0}(D,\omega)=C^{s=0}(D,\omega)\sum_l N_{l\,S}\left|{\cal A}^{s=0}_{l,S}
\right|^2\,,\label{sigmas0}
\ee
where the normalization factor is
\be
C^{s=0}(D,\omega)=\left(\frac{2\pi}{\omega}\right)^{D-2}\frac{1}{\Omega_{D-2}}
=\frac{(4\pi)^{(D-3)/2}\Gamma[(D-1)/2]}{\omega^{D-2}}\,,
\label{C0}
\ee
and the multiplicities of the scalar spherical harmonics are
\be
N_{l\,S}=\frac{(2l+D-3)(l+D-4)!}{l!(D-3)!}\,.
\label{N0}
\ee
Equation (\ref{C0}) follows from the decomposition of a plane wave in spherical waves. In four
dimensions, Eq.\ (\ref{N0}) gives the well-known result $N_{l\,S}=2l+1$.

Spherical harmonics and partial wave expansion are different for vector and tensor fields. This leads to
different expressions for the normalization and multiplicity factors. (This fact has been overlooked in the
literature. See, for instance, Ref.\ \cite{Kanti:2004nr}.)

Consider a vector field $V^\mu$ in the transverse gauge. The decomposition of a transverse  plane
wave in spherical waves gives the factor (see the Appendix \ref{crossapp} and Ref.\
\cite{Crispino:2000jx})
\be C^{s=1}(D,\omega)=\frac{C^{s=0}(D,\omega)}{D-2}\,.
\label{Cs1}
\ee
Following KI, we first decompose $V^\mu$ in the divergence of a scalar field plus a divergenceless
vector. These components are then expanded in scalar and vector spherical harmonics. The
multiplicities of the scalar harmonics are given by Eq.\ (\ref{N0}). The multiplicities of the
vector harmonics are
\be
N_{l\,V}=\frac{l(l+D-3)(2l+D-3)(l+D-5)!}{(l+1)!(D-4)!}\,.
\label{N1}
\ee
The cross section is
\be
\sigma^{s=1}=C^{s=1}(D,\omega)\sum_l\left[N_{l\,S} \left|{\cal A}^{s=1}_{l\,S}\right|^2+
N_{l\,V}\left|{\cal A}^{s=1}_{l\,V}\right|^2\right]\,.
\ee
Finally, let us consider the graviton field. The decomposition of a plane wave in spherical waves
yields the normalization factor (see the Appendix \ref{crossapp} for details)
\be
C^{s=2}(D,\omega)=\frac{2}{D(D-3)}\,C^{s=0}(D,\omega)\,.
\label{relCs}
\ee
The gravitational perturbation is first decomposed in scalar, vector and tensor components on the
tangent space of $S^{D-2}$. These components are then expanded in scalar, vector and tensor
harmonics. The corresponding multiplicities are $N_{l\,S}$, $N_{l\,V}$ and \cite{Chodos:1983zi}
\be N_{l\,T}=\frac{1}{2}\frac{(D-1)(D-4)(l+D-2)(l-1)(2l+D-3) (l+D-5)!}{(l+1)!(D-3)!}
\label{N2}\,,
\ee
respectively. The graviton absorption cross section is
\be \sigma^{s=2}=C^{s=2}(D,\omega)\sum_l \left[ N_{l\,S}\left|{\cal A}^{s=2}_{l\,S}\right|^2+
N_{l\,V}\left|{\cal A}^{s=2}_{l\,V}\right|^2 +N_{l\,T}\left|{\cal A}^{s=2}_{l\,T}\right|^2\right]\,.
\label{crossgrav}
\ee
Note that in $D=4$, accidentally, the conversion factors do not depend on the spin. The
four-dimensional cross sections read
\be
\sigma^{s}(4,\omega)=\frac{\pi}{\omega^2}\sum_l(2l+1)|{\cal A}^{s}_{l}|^2\,,\qquad s=0,1,2\,.
\ee
(See also the discussion in Section \ref{sec:highen} below.)
\section{\label{sec:lowen}Low-energy absorption probabilities}
Since low frequencies ($\omega\ll 1$) give a substantial contribution to the total power emission,
it is worthwhile to derive an analytical expression for this limiting case. The low-energy
approximation uses a matching procedure to find a solution on the whole spacetime for any value of
$p$. Let us define the near-horizon region $r-1\ll 1/\omega$ and the asymptotic region $r\gg 1$. To
solve the wave equation in the near-horizon region, we set
\be
\Psi=r^{\beta^0/2}~\Xi\,,
\ee
where $\beta^0=(D-2)(p+1)$. With this substitution, the wave equation (\ref{eveq}) is cast in the
form
\be \frac{f}{r^{\beta^0}}\partial_r \left (fr^{\beta^0}\partial_r
\Xi\right )+ {\Biggl [} \omega^2 +\frac{f}{r^2}\left
(-l(l+D-3)+\frac{p(D-3)(D-2)}{2}+\frac{p^2(D-2)^2}{4}\right ){\Biggl
]}\Xi=0 \,.
\label{nr}
\ee
Changing variables to
\be v=1-\frac{1}{r^{D-3}}\,,
\ee
Eq.\ (\ref{nr}) reads
\be
(1-v)^2v^2(3-D)^2\partial^2_v\Xi-(3-D)v(1-v)^2\left
(\frac{p(D-2)v}{(1-v)}-(3-D)\right )\partial_v\Xi+\left ((\omega
r)^2+va \right )\Xi=0\,,
\ee
where
\be
a=-l(l+D-3)+\frac{p(D-3)(D-2)}{2}+\frac{p^2(D-2)^2}{4}\,.
\ee
This equation can be put in a standard hypergeometric form by defining
\be
p_h=\frac{p(D-2)}{D-3}\,,\quad \omega_h=\frac{\omega}{D-3}\,,\quad
a_h=\frac{a}{(D-3)^2}\,,
\ee
and setting
\beq
\Xi&=&v^{c_1}(1-v)^{c_2}{\cal F}\,,\\ c_1&=&-i\omega_h\,,\\
c_2&=&\frac{1}{2}\left (1+p_h-b \right
)=\frac{1}{2}\frac{\beta^0-1-(D-3)b}{D-3} \,,\\
b&=&\sqrt{(1+p_h)^2-4a_h-4\omega_h^2}=\frac{1}{D-3}\sqrt{(D+2l-3)^2-4\omega^2}\,.
\eeq
The result is
\be
v(1-v)\partial^2_v {\cal F}+\left (\gamma-v(1+\alpha+\beta) \right
)\partial_v{\cal F}-\alpha\beta {\cal F}=0\,,
\label{hypergeom}
\ee
where
\beq
\gamma=1-2i\omega_h\,,\quad \alpha=\frac{1}{2}\left
(1-p_h-2i\omega_h-b\right )\,,\quad \beta=\frac{1}{2}\left
(1+p_h-2i\omega_h-b\right )\,.
\eeq
The most general solution of Eq.\ (\ref{hypergeom}) in the neighborhood of $v=0$ is
\be
{\cal F}=A\,F(\alpha,\beta,\gamma,v)+
B\,v^{1-\gamma} F(\alpha-\gamma+1,\beta-\gamma+1,2-\gamma,v)\,.
\ee
Since the second term describes an outgoing wave near the horizon, we set $B=0$. The asymptotic
behavior of the near-horizon solution is
\be
\Xi \sim r^{-2i\omega-(\beta^0-1)-(D-3)b/2} \times
\frac{\Gamma(\gamma)\Gamma(\alpha+\beta-\gamma)}{\Gamma(\alpha)\Gamma(\beta)}
+r^{-(\beta^0-1)+(D-3)b/2}\frac{\Gamma(\gamma)\Gamma(\gamma-\alpha-\beta)}{\Gamma(\gamma-\alpha)
\Gamma(\gamma-\beta)}\,,
\label{match1}
\ee
where we have used the property of the hypergeometric functions:
\begin{eqnarray}
& \hspace{-2cm} F(\alpha,\beta,\gamma,v)=
(1\!-\!v)^{\gamma-\alpha-\beta} \times
\frac{\Gamma(\gamma)\Gamma(\alpha+\beta-\gamma)}{\Gamma(\alpha)\Gamma(\beta)}
\,F(\gamma-\alpha,\gamma-\beta,\gamma\!-\!\alpha\!-\!\beta\!+\!1,1\!-\!v)
& \nonumber \\ & \hspace{-0.2cm}+
\frac{\Gamma(\gamma)\Gamma(\gamma-\alpha-\beta)}{\Gamma(\gamma-\alpha)\Gamma(\gamma-\beta)}
\,F(\alpha,\beta,-\gamma\!+\!\alpha\!+\!\beta\!+\!1,1\!-\!v)\,.
\label{transformation law}
\end{eqnarray}
In the asymptotic region, Eq.\ (\ref{nr}) can be written as
\be \partial^2_r \Xi +\frac{\beta^0}{r}\partial_r \Xi+ {\Biggl [}
\omega^2 +\frac{f}{r^2}\left
(-l(l+D-3)+\frac{p(D-3)(D-2)}{2}+\frac{p^2(D-2)^2}{4}\right ){\Biggl
]}\Xi=0 \,.
\label{nr2}
\ee
The solution of this equation is
\be
\Xi = C_1r^{(1-\beta^0)/2}J_{{\hat b}/2}(\omega r)+C_2r^{(1-\beta^0)/2}Y_{{\hat b}/2}(\omega r)\,,
\ee
where ${\hat b}=\sqrt{(1-\beta^0)^2-4a}$. Expanding $\Xi$ for small $\omega r$, we obtain
\be
\Xi \sim C_1 \frac{(\omega/2)^{{\hat b}/2}}{\Gamma(1+\hat b/2)}r^{(1-\beta^0+{\hat
b})/2}-C_2\frac{(2/\omega)^{{\hat b}/2}\Gamma(\hat b/2)}{\pi}r^{(1-\beta^0-{\hat b})/2}\,.
\label{match2}
\ee
Equation (\ref{match2}) is matched to Eq.\ (\ref{match1}). As $(D-3)b \sim {\hat b}$ for
$\omega\ll 1$, we find
\be
\frac{\Gamma(\gamma-\alpha-\beta)\Gamma(\alpha)\Gamma(\beta)}
{\Gamma(\alpha+\beta-\gamma)\Gamma(\gamma-\alpha)\Gamma(\gamma-\beta)}=-\frac{C_1}{C_2}\frac{\pi
(\frac{\omega}{2})^{\hat b}} { \Gamma(1+\frac{{\hat b}}{2})\Gamma(\frac{{\hat b}}{2})}\,, \ee
\be \frac{C_1}{C_2}=-\left (\frac{2}{\omega}\right
)^{D+2l-3}\frac{\Gamma\left(l+\frac{D-3}{2}\right)^2\left(l+\frac{D-3}{2} \right
)\Gamma(\gamma-\alpha-\beta)\Gamma(\alpha)\Gamma(\beta)}
{\pi\Gamma(\alpha+\beta-\gamma)\Gamma(\gamma-\alpha)\Gamma(\gamma-\beta)}\,, \ee
\be \frac{C_1}{C_2}=-\left (\frac{2}{\omega}\right
)^{D+2l-3}\frac{\Gamma\left(l+\frac{D-3}{2}\right)^2\left(l+\frac{D-3}{2} \right )\Gamma\left
(1+\frac{2l}{D-3}\right )^2\Gamma(\beta)^2\left ( 1+\frac{2l}{D-3}\right )\sin\pi\beta\sin\frac{2l\pi}{D-3}}
{\pi^2 \alpha^2\sin{\pi\alpha}\Gamma(-\alpha)^2}\,. \ee
It is straightforward to show that the reflection coefficient is
\be {\cal R}=\frac{R}{T} =\frac{C_1-iC_2}{C_1+iC_2}\,. \ee
Therefore, the absorption probability in the $\omega\ll 1$ approximation is
\be
\left|{\cal A}\right|^2=1-|{\cal R}|^2=4\pi\left (
\frac{\omega}{2}\right )^{D+2l-2} \frac{\Gamma\left
(1+\frac{2l+p(D-2)}{2(D-3)}\right )^2\Gamma\left
(1+\frac{2l-p(D-2)}{2(D-3)}\right )^2}{\Gamma\left
(1+\frac{2l}{D-3}\right )^2\Gamma\left (l+\frac{(D-1)}{2}\right )^2}
\label{AbsCoeff}\,.
\ee
The result for the scalar waves \cite{Kanti:2002nr} are obtained by setting $p=0$
and $l=0\,,1\dots$. Setting $p=0$ (2) and $l=2\,,3\dots$ we obtain the low-energy absorption
probability for the gravitational tensor (vector) perturbations. The gravitational scalar
perturbation cannot be dealt analytically. Numerical results give
\be
p_{\rm grav\, scalar}\sim 2+0.674D^{-0.5445}\,.
\ee
Using Eqs.\ (\ref{N0}), (\ref{N1}), (\ref{relCs}), (\ref{N2}) and (\ref{AbsCoeff}) it is
straightforward to compute the low-energy absorption cross section for the gravitational waves. For
instance, recalling that the four-dimensional absorption probabilities of scalar and vector
perturbations are equal \cite{Chandrasekar} and the tensor contribution vanishes, the contribution of
the $l=2$ mode in four dimensions is
\be \sigma_l= \frac{4\pi}{45}\omega^4\,. \ee
This result agrees with the well-known result of Ref.\ \cite{Page:1976df}.
\section{\label{sec:highen}High-energy absorption cross sections}
The wave equation can also be solved in the high-energy limit, where the absorption cross section
is expected to approximate the cross section for particle capture. This has been verified in
a number of papers for the scalar field. (See Ref.~\cite{Kanti:2004nr} and references therein.) We
will now extend this result to spin-1 and -2 fields in any dimension $D$. In the high-$l$ limit,
the multiplicities satisfy the relations
\begin{eqnarray}
N_{l\,,S}=\frac{1}{D-2}\left[N_{l\,,S}+N_{l\,,V}\right]
=\frac{2}{D(D-3)}\left[N_{l\,,S}+N_{l\,,V}+N_{l\,,T}\right]\,,\quad l\rightarrow \infty
\,,\quad D>2\,.
\label{coinc}
\end{eqnarray}
The above relations also hold for any $l$ when $D=4$. In that case, Chandrasekhar
\cite{Chandrasekar} showed that ${\cal A}_{l\,S}={\cal A}_{l\,V}$ for any $l$ and
$N_{l\,T}(4,2)=0$. Therefore, in four dimensions the cross section does not depend on the particle
spin.

Since high frequencies can easily penetrate the gravitational potential barrier, the absorption
probabilities of all fields, $|{\cal A}(D,l,\omega)|^2$, tend to 1 as $\omega \rightarrow \infty$.
A WKB analysis shows that the absorption probability is nonzero for $l\lesssim\omega$. Therefore,
the cross section in the high-energy limit must include the contribution from all $l\lesssim
\omega$. The largest contribution to the cross section when $\omega \rightarrow \infty$ is given by
high-$l$ modes. Since the high-$l$ limit of the wave equation is independent of the kind of
perturbation (see Section \ref{sec:metric}), the wave equation takes a universal form in this
limit, i.e.\ ${\cal A}_{l\,S}={\cal A}_{l\,V}={\cal A}_{l\,T}$. From Eqs.\
(\ref{Cs1})-(\ref{crossgrav}) and (\ref{coinc}) it follows that the absorption cross sections for
spin-0, -1 and -2 fields is equal in the limit $\omega\rightarrow \infty$.
\section{\label{sec:final}Total energy emission}
The total energy flux for gravitational waves is
\be
\frac{dE}{dt}=\frac{dE_S}{dt}+\frac{dE_V}{dt}+\frac{dE_T}{dt}=\sum_l \int \frac{d \omega}{2\pi}
\frac{\omega}{e^{\omega/T_H}-1}\left (N_{l\,S} |{\cal A}_{l\,S}^{s=2}|^2+N_{l\,V} |{\cal
A}_{l\,V}^{s=2}|^2+N_{l\,T} |{\cal A}_{l\,T}^{s=2}|^2 \right )\,,
\label{totpower}
\ee
where the Hawking temperature is $T_H=(D-3)/(4\pi)$ and the counting of helicities is included in
the multiplicity factors. The total absorption probabilities can be computed numerically.
Equation (\ref{eveq}) is integrated from a point near the horizon (typically $r-1 \sim 10^{-6}$),
where the field behavior is given by Eq.~(\ref{def-bcrmais}). The numerical result is then
compared to Eq.~(\ref{def-bc}) at large $r$. A better accuracy is achieved by considering the
next-to-leading order correction terms \cite{Berti:2003si}
\be
\Psi(r)\rightarrow T \left (1+\frac{\varepsilon}{r} \right ) e^{-i\omega r_*}+ R\left
(1-\frac{\varepsilon}{r}\right ) e^{i\omega r_*}\,,\quad r_*\rightarrow \infty\,,
\label{def-bc2}
\ee
where
\be
\varepsilon=-i\frac{l(l+D-3)+(D-2)(D-4)/4}{2\omega}\,.
\ee
The emission rates and the total integrated power for various fields are summarized in Tables
\ref{tab:totalpower}-\ref{tab:emission rates}. For sake of comparison with previous works, the
values for lower-spin fields are taken from Ref.\ \cite{Kanti:2004nr}. (We checked these results
with our numerical codes and found agreement within numerical uncertainties.) The results in the
tables are normalized to the four-dimensional values. In four dimensions, the radiated power
${\cal P}$ is
\be {\cal P}^{s=0}=2.9\times 10^{-4}\,r_+^{-2}\,,\quad {\cal P}^{s=1/2}=1.6\times 10^{-4}\,r_+^{-2} \,,\quad
{\cal P}^{s=1}=6.7\times 10^{-5}\,r_+^{-2} \,,\quad {\cal P}^{s=2}=1.5\times 10^{-5}\, r_+^{-2} \,. \ee
The spin-0, -1/2 and -1 values are per d.o.f., whereas the graviton value includes the
contribution of the two helicities \footnote{Since the number of graviton d.o.f.\ (helicities)
depends on the spacetime dimensionality $D$, here and throughout the paper we give the {\it total}
rate and the {\it total} power for the graviton field, rather than the rate and power {\it per
d.o.f.}}. The four-dimensional emission rates are
\be
{\cal R}^{s=0}=1.4\times 10^{-3}\,r_+^{-1}\,, \quad
{\cal R}^{s=1/2}=4.8\times 10^{-4}\,r_+^{-1}\,, \quad
{\cal R}^{s=1}=1.5\times 10^{-4}\,r_+ ^{-1}\,,\quad
{\cal R}^{s=2}=2.2\times 10^{-5}\, r_+ ^{-1}\,.
\ee
\begin{table}[h]
\caption{\label{tab:totalpower} Total radiated power ${\cal P}$ into different channels. The first three rows
correspond to fields propagating on the brane. The last row gives the power radiated in bulk gravitons
normalized to the four-dimensional case.}
\begin{ruledtabular}
\begin{tabular}{ccccccccc}  \hline
$D$ &4&5&6&7&8&9&10&11\\ \hline
{\rm Scalars}&1 &8.94 &36&99.8 &222&429&749&1220\\ 
{\rm Fermions}&1 &14.2 &59.5&162 &352&664&1140&1830\\ 
{\rm Gauge Bosons}&1 &27.1 &144&441 &1020&2000&3530&5740\\ 
{\rm Gravitons}&1 &103 &1036& 5121 &$2\times 10^4$&$7.1\times
10^{4}$&$2.5\times 10^5$&$8\times 10^5$\\ 
\end{tabular}
\end{ruledtabular}
\end{table}
The above values for fermions, bosons and graviton agree with Page's results (See Table
I in Ref.\ \cite{Page:1976df}). Some features of the numerical results of Table \ref{tab:totalpower} are
worth discussing:
\begin{itemize}
\item The relative contributions of the higher partial waves increase with $D$. For instance, in
four dimensions the contribution of the $l=2$ mode is two orders of magnitude larger than the
contribution of the $l=3$ mode. The $l=2$ and $l=3$ contributions are roughly equal in $D=8$.
More energy is channeled in $l=3$ mode than in $l=2$ mode for $D\ge 9$. (The largest tensor
contribution in ten dimensions comes from the $l=4$ mode.) Contributions from high $l$ are needed
to obtain accurate results for large $D$. For instance, in ten dimensions the first 10 modes must
be considered for a meaningful result. Therefore, precise values for very large $D$ require the
most CPU-time. The values in Table \ref{tab:totalpower} have a 5\% accuracy.
\item  The total power radiated in gravitons increases more rapidly with $D$ than the power
radiated in lower-spin fields. This is due to the increase in the multiplicity of the tensor
perturbations, which is larger than the scalar multiplicity by a factor $D^2$ at high $D$.
Therefore, the main contribution to the total power comes from tensor (and vector) modes. For
instance, in ten dimensions the tensor modes contribute roughly half of the total power output.
\end{itemize}
Table \ref{tab:totalpower3} gives the fraction of radiated power per d.o.f.\ normalized to the
scalar field. In four dimensions, the graviton channel is only about 5\% of the scalar channel.
Therefore, the power loss in gravitons is negligible compared to the power loss in lower-spin
fields. This conclusion is reversed in higher dimensions. For instance, the graviton loss is about
35 times higher than the scalar loss in $D=11$. Graviton emission is expected to dominate the black
hole evaporation at very high $D$.
\begin{table}[h]
\caption{\label{tab:totalpower3} Fraction of radiated power per d.o.f.\ normalized to the scalar field. The
graviton d.o.f.\ (number of helicity states) are included in the results.}
\begin{ruledtabular}
\begin{tabular}{ccccccccc}  \hline
$D$ &4&5&6&7&8&9&10&11\\ \hline
{\rm Scalars}        &1      &1    &1    &1   &1   &1  &1&1\\ 
{\rm Fermions}       &0.55   &0.87   &0.91   &0.89    &0.87  &0.85  &0.84&0.82\\ 
{\rm Gauge Bosons}&0.23    &0.69   &0.91   &1.0  &1.04  &1.06  &1.06&1.07\\ 
{\rm Gravitons }      &0.053   &0.61    &1.5    &2.7   &4.8   &8.8   &17.7 &34.7\\ 
\end{tabular}
\end{ruledtabular}
\end{table}
The particle emission rates per d.o.f.\ are shown in Table III. The relative emission rates of
different fields can be obtained by summing on the brane d.o.f.\ For instance, the relative
emission rates of standard model charged leptons (12 d.o.f.) and a 11-dimensional bulk graviton are
roughly 1:1. This ratio becomes $\sim$ 40:1 in five dimensions.
\begin{table}[h]
\caption{\label{tab:emission rates} Fraction of emission rates per d.o.f.\ normalized to the scalar field. The
graviton result includes all the helicity states and counts as one d.o.f.}
\begin{ruledtabular}
\begin{tabular}{ccccccccc}  \hline
$D$ &4&5&6&7&8&9&10&11\\ \hline
{\rm Scalars}        &1      &1    &1    &1   &1   &1  &1&1\\ 
{\rm Fermions}       &0.37   &0.7   &0.77   &0.78    &0.76  &0.74  &0.73&0.71\\ 
{\rm Gauge Bosons}&0.11    &0.45   &0.69   &0.83  &0.91  &0.96  &0.99&1.01\\ 
{\rm Gravitons }    &0.02    &0.2   &0.6  &0.91   &1.9   &2.5 &5.1  &7.6\\ 
\end{tabular}
\end{ruledtabular}
\end{table}
To illustrate the relevance of these results for black hole in particle colliders, let us consider the
minimal $U(1)\times SU(2)\times SU(3)$ standard model with three families and one Higgs field on a thin
brane with fundamental Planck scale = 1 TeV. For black holes with mass $\sim$ few TeV the Hawking
temperature is generally above $100$ GeV. The temperature of a six-dimensional black hole with mass
equal to 5 (100) TeV is $\sim 282$ (133) GeV. The temperature increases with the spacetime dimension at
fixed mass. Therefore, all d.o.f.\ can be considered massless. (Considering massive gauge bosons does
not affect the conclusions significantly.) The spin-0, -1/2 and -1 d.o.f.\ on the brane are 4 (complex
Higgs doublet), 90 (quarks + charged leptons + neutrinos) and 24 (massless gauge bosons), respectively.
The relative emissivities for this model are shown in Table \ref{tab:totalpowerSM}. Although the
graviton emission is highly enhanced, the large number of brane d.o.f.\ implies that the brane channel
dominates on the bulk channel. However, power loss in the bulk is significant and cannot be neglected
at high $D$; about 1/4 of the initial black hole mass is lost in the 11-dimensional bulk. This implies
a larger-than-expected missing energy in particle colliders.
\begin{table}[h]
\caption{\label{tab:totalpowerSM} Percentage of power going into each field species for the minimal $U(1)\times
SU(2)\times SU(3)$ standard model with three families and one Higgs field above the spontaneous symmetry
breaking scale. The four-dimensional results are taken from Ref.\ \cite{Page:1976df} and the higher dimensional
results for fermions and gauge fields are taken from Ref.\ \cite{Kanti:2004nr}.}
\begin{ruledtabular}
\begin{tabular}{ccccccccc}  \hline
$D$ &4&5&6&7&8&9&10&11\\ \hline
{\rm Scalars}        &6.8     &4.0    &3.7    &3.6     &3.6   &3.5 &3.3   &2.9\\ 
{\rm Fermions}       &83.8   &78.7   &75.0   &72.3    &69.9   &66.6 &61.6  &53.4\\ 
{\rm Gauge Bosons}   &9.3     &16.7   &20.0   &21.7    &22.3   &22.2 &20.7   &18.6\\ 
{\rm Gravitons }      &0.1   &0.6    &1.3    &2.4      &4.2     &7.7 &14.4    &25.1\\ 
\end{tabular}
\end{ruledtabular}
\end{table}
%
%
\section{\label{sec:concl}Conclusions}
In this paper we have computed the absorption cross section and the total power carried by gravitons in
the evaporation process of a higher-dimensional non-rotating black hole. We find that the power loss in
the graviton channel is highly enhanced in higher-dimensional spacetimes. This has important
consequences for the detection of microscopic black hole formation in particle colliders and
ultrahigh-energy cosmic ray observatories, where a larger bulk emission implies larger missing energies
and lower multiplicity in the visible channels. Despite the increase in graviton emissivity, for
$4<D\le 11$ a non-rotating black hole in the Schwarzschild phase will emit mostly on the brane due to
the higher number of brane d.o.f. \footnote{This conclusion holds only if the particle content at TeV
energies is mostly made of fields propagating on the brane. If non-standard model d.o.f.\ open up at
the TeV scale, such as vector multiplets propagating in the bulk, black hole emission could occur
mostly in the bulk.} However, black hole energy loss in the bulk cannot be neglected in presence of
extra dimensions. Graviton emission is expected to dominate the black hole evaporation at very high
$D$.
\section*{Acknowledgements}
We are very grateful to De-Chang Dai for pointing out some typos in previous versions. The authors thank
Akihiro Ishibashi for useful correspondence. VC acknowledges financial support from FCT through the PRAXIS XXI
program, and from the Funda\c c\~ao Calouste Gulbenkian through the Programa Gulbenkian de Est\'{\i}mulo \`a
Investiga\c c\~ao Cient\'{\i}fica.
\appendix*
\section{Derivation of the cross sections}
\label{crossapp}
In this appendix we derive in detail the cross sections for spin-0, -1 and -2 fields in generic
dimensions. The harmonic expansion is performed on the spherical manifold $S^n$ with
metric $\gamma_{ab}$ and coordinates $y^a$ ($a=1,\dots,n=D-2$). We also define the set of coordinates
$x^i=(r,y^a)$. In the limit $r\rightarrow\infty$ the wave vector of a massless particle is
\be
k^\mu=(\omega,\vec k)\,,
\ee
where $\vec k=(k^i)=\omega\hat k$. The direction of the plane wave is given by the
$(n+1)$-dimensional unit vector $\hat k$.
\subsection*{Scalar perturbations}
Let us consider an incoming scalar plane wave $\phi^{(\hat k,\omega)}$ with unit flux and wave
vector $\vec k=\omega\hat k$ at infinity:
\be
\phi^{(\hat k,\omega)}
\rightarrow e^{-{\rm i}\omega t-{\rm i}\vec k\vec x}\,.
\ee
This wave can be expanded in spherical waves $\Psi^{(\lambda,\omega)}$ parametrized by a discrete
index $\lambda$:
\be
\phi^{(\hat k,\omega)}=\sum_\lambda\alpha(\hat k,\omega;\lambda)
\Psi^{(\lambda,\omega)}\,.
\ee
In general, $\lambda=(l,m)$. However, we leave the index implicit to simplify the generalization to
higher spins. The asymptotic behavior of the fields $\Psi$ is
\be
\Psi^{(\lambda,\omega)}\rightarrow \frac{e^{-{\rm i}\omega t-{\rm i}\omega r}}{r^{n/2}}
Y^{(\lambda)}\,,
\ee
where the spherical harmonics, $Y^{(\lambda)}$, satify the normalization and orthogonality
conditions
\be
\int d\Omega_nY^{(\lambda)}Y^{(\lambda')}=\delta^{\lambda\lambda'}\,.
\ee
From the above equations it follows
\be
r^n\int d\Omega_n \Psi^{*(\lambda,\omega)}\Psi^{(\lambda,\omega)}=1\,,
\ee
i.e.\ the wave $\Psi^{(\lambda,\omega)}$ describes one ingoing particle per unit time. If ${\cal
A}^{s=0}(\lambda,\omega)$ is the absorption coefficient for a spherical wave with angular profile
$Y^{(\lambda)}$, the cross section associated to the plane wave $(\hat k,\omega)$ is
\be
\sigma^{\hat k,\omega}=\sum_\lambda|\alpha(\hat k,\omega;\lambda)|^2
|{\cal A}^{s=0}(\omega,\lambda)|^2\,.
\ee
The average on the wave direction of the initial state is obtained by integrating $\hat k$ over the
sphere and dividing by the unit volume $\Omega_n$:
\be
\sigma=\frac{1}{\Omega_n}\sum_\lambda\int d\Omega_n^{(\hat k)}
|\alpha(\hat k,\omega;\lambda)|^2|{\cal A}^{s=0}(\omega,\lambda)|^2\,.
\ee
From the previous equation, we obtain
\be
C^{s=0}(n+2,\omega)=\frac{1}{\Omega_n}\int d\Omega_n^{(\hat k)}
|\alpha(\hat k,\omega;\lambda)|^2\,.
\ee
The coefficients $\alpha$ are given by the coefficients of the expansion
\be
\tilde\phi^{(\hat k,\omega)}=\sum_\lambda\alpha(\hat k,\omega;\lambda)
\tilde\Psi^{(\lambda,\omega)}\,,
\ee
where
\be
\tilde\Psi^{(\lambda,\omega)}\equiv\frac{e^{-{\rm i}\omega t-
{\rm i}\omega r}}{r^{n/2}}Y^{(\lambda)}\,,\qquad
\tilde\phi^{(\hat k,\omega)}\equiv e^{-{\rm i}\omega t-{\rm i}
\vec k\vec x}\,,
\ee
are defined in the flat $(n+1)$-dimensional Euclidean space. From
\be
\int d^{n+1}x\tilde\Psi^{*(\lambda',\omega')}\tilde\Psi^{(\lambda,\omega)}
=2\pi\delta(\omega-\omega')\delta_{\lambda\lambda'}
\ee
it follows
\be
(2\pi)^2\delta(\omega-\omega')\delta(\omega'-
\omega'')|\alpha(\hat k,\omega;\lambda)|^2
=\int d^{n+1}xd^{n+1}x'\tilde\Psi^{*(\lambda,\omega)}(x)
\tilde\phi^{(\hat k,\omega')}(x)\tilde\phi^{*(\hat k,\omega')}(x')
\tilde\Psi^{(\lambda,\omega'')}(x')\,.
\ee
Multiplying by $\omega^{\prime n}$ and integrating in $d\Omega_n^{(\hat k)}d\omega'$, we obtain
\beq
&&\omega^n(2\pi)^2\delta(\omega-\omega'')\int d\Omega^{(\hat k)}_n
|\alpha(\hat k,\omega;\lambda)|^2\nonumber\\
&=&\int d^{n+1}xd^{n+1}x'\tilde\Psi^{*(\lambda,\omega)}(x)
\tilde\Psi^{(\lambda,\omega'')}(x')\left[\int d\omega'\omega^{\prime n}
\int d\Omega^{(\hat k)}_ne^{-{\rm i}\vec k(\vec x-\vec x')}\right]\nonumber\\
&=&(2\pi)^{n+1}\int d^{n+1}xd^{n+1}x'\delta(\vec x-\vec x')
\tilde\Psi^{*(\lambda,\omega)}(x)\tilde\Psi^{(\lambda,\omega'')}(x')
=(2\pi)^{n+2}\delta(\omega-\omega'')\,.
\nonumber\\
\eeq
Finally, the normalization factor is
\be
C^{s=0}(n+2,\omega)=\frac{1}{\Omega_n}\int d\Omega_n^{(\hat k)}
|\alpha(\hat k,\omega;\lambda)|^2=\frac{1}{\Omega_n}\left(\frac{2\pi}{\omega}\right)^n\,.
\ee
%
\subsection*{Vector perturbations}
Let us consider an incoming transverse vector plane wave $\phi_\mu^{(p,\hat k,\omega)}$ with unit
flux and wave vector at infinity:
\be
\phi_\mu^{(p,\hat k,\omega)}
\rightarrow e^{-{\rm i}\omega t-{\rm i}\vec k\vec x}
\left(\begin{array}{c} 0 \\ \epsilon^p_i(\hat k) \\ \end{array} \right)\,,
\ee
where $\epsilon^p_i(\hat k)$ are the transverse polarization vectors ($p=1,\dots,n$) and
\be
\hat k^i\epsilon^p_i=0\,,\qquad
\epsilon^p_i\epsilon^{p'}_j\delta^{ij}=\delta^{pp'}\,,\qquad
\sum_{p=1}^n\epsilon^p_i\epsilon^p_j=\delta_{ij}+\hbox{ terms in }k_i,k_j\,.
\ee
We assume $x^i$ to be asymptotically Euclidean coordinates for simplicity. The wave can be
expanded in spherical waves $\Psi^{(\lambda,\omega)}$ parametrized by a discrete index $\lambda$,
which includes the polarizations:
\be
\phi_\mu^{(p,\hat k,\omega)}=\sum_\lambda\alpha(p,\hat k,\omega;\lambda)
\Psi_\mu^{(\lambda,\omega)}\,.
\ee
The asymptotic fields $\Psi_\mu$ are
\be
\Psi_\mu^{(\lambda,\omega)}\rightarrow
\frac{e^{-{\rm i}\omega t-{\rm i}\omega r}}{r^{n/2}}
\left(\begin{array}{c} 0 \\ 0 \\  rY^{(\lambda)}_a \\ \end{array}\right)\,,
\ee
where $Y^{(\lambda)}_a$ satisfy the normalization and orthogonality conditions
\be
\int
d\Omega_nY_{a}^{(\lambda)}Y_{b}^{(\lambda')}\gamma^{ab}=\delta^{\lambda\lambda'}\,.
\ee
From the above equation it follows
\be
r^n\int d\Omega_n \Psi_\mu^{*(\lambda,\omega)}\Psi_\mu^{(\lambda,\omega)}g^{\mu\nu}=1\,,
\ee
i.e.\ a wave $\Psi_\mu^{(\lambda,\omega)}$ describes one ingoing particle per unit time.  If ${\cal
A}^{s=1}(\lambda,\omega)$ is the absorption coefficient for a spherical wave with angular profile
$Y_a^{(\lambda)}$, the cross section associated to the plane wave $(p,\hat k,\omega)$ is
\be
\sigma^{p,\hat k,\omega}=\sum_\lambda|\alpha(p,\hat k,\omega;\lambda)|^2
|{\cal A}^{s=1}(\omega,\lambda)|^2\,.
\ee
The average on the wave direction and polarization of the initial state is obtained by summing on
$p$, integrating $\hat k$ over the sphere and dividing by the unit volume $\Omega_n$ and the number
of polarizations $n$:
\be
\sigma=\frac{1}{n\Omega_n}\sum_\lambda\int d\Omega_n^{(\hat k)}
\sum_{p=1}^n|\alpha(p,\hat k,\omega;\lambda)|^2|{\cal A}^{(\omega,\lambda)}|^2\,.
\ee
From the previous equation, we obtain
\be
C^{s=1}(n+2,\omega)=\frac{1}{n\Omega_n}\int d\Omega_n^{(\hat k)}\sum_{p=1}^n
|\alpha(p,\hat k,\omega;\lambda)|^2\,.
\ee
The coefficients $\alpha$ are the coefficients of the expansion
\be
\tilde\phi_\mu^{(p,\hat k,\omega)}=\sum_\lambda\alpha(p,\hat k,\omega;\lambda)
\tilde\Psi_\mu^{(\lambda,\omega)}\,,
\ee
where
\be
\tilde\Psi_i^{(\lambda,\omega)}\equiv
\frac{e^{-{\rm i}\omega t-{\rm i}\omega r}}{r^{n/2}}
\left(\begin{array}{c} 0 \\  rY^{(\lambda)}_a \\ \end{array}\right)\,,\qquad
\tilde\phi_i^{(p,\hat k,\omega)}\equiv e^{-{\rm i}\omega t-{\rm i}\vec k\vec x}
\epsilon^p_i(\hat k)\,,
\ee
are defined in the flat $(n+1)$-dimensional Euclidean space. From
\be
\int d^{n+1}x\tilde\Psi_i^{*(\lambda',\omega')}\tilde\Psi^{i(\lambda,\omega)}
=2\pi\delta(\omega-\omega')\delta_{\lambda\lambda'}
\ee
it follows
\be
(2\pi)^2\delta(\omega-\omega')\delta(\omega'-
\omega'')|\alpha(p,\hat k,\omega;\lambda)|^2
=\int d^{n+1}xd^{n+1}x'\tilde\Psi_i^{*(\lambda,\omega)}(x)
\tilde\phi^{i(p,\hat k,\omega')}(x)\tilde\phi_j^{*(p,\hat k,\omega')}(x')
\tilde\Psi^{j(\lambda,\omega'')}(x')\,.
\ee
Multiplying by $\omega^{\prime n}$, summing over $p$ and integrating in $d\Omega_n^{(\hat
k)}d\omega'$, we obtain
\beq
&&\omega^n(2\pi)^2\delta(\omega-\omega'')\int d\Omega^{(\hat k)}_n
\sum_{p=1}^n|\alpha(p,\hat k,\omega;\lambda)|^2\nonumber\\
&=&\int d^{n+1}xd^{n+1}x'\tilde\Psi_i^{*(\lambda,\omega)}(x)
\tilde\Psi^{j(\lambda,\omega'')}(x')\left[\int d\omega'\omega^{\prime n}
\int d\Omega^{(\hat k)}_n\left(\sum_{p=1}^n\epsilon^{i(p)}(\hat k)
\epsilon^{*(p)}_j(\hat k)\right)
e^{-{\rm i}\vec k(\vec x-\vec x')}\right]\nonumber\\
&=&(2\pi)^{n+1}\int d^{n+1}xd^{n+1}x'\delta(\vec x-\vec x')
\tilde\Psi_i^{*(\lambda,\omega)}(x)\tilde\Psi^{j(\lambda,\omega'')}(x')
\left(\delta^i_j+\hbox{ terms in }k^i,k_j\right)
=(2\pi)^{n+2}\delta(\omega-\omega'')\,.
\nonumber\\
\eeq
Finally, the normalization factor is
\be
C^{s=1}(n+2,\omega)=\frac{1}{n\Omega_n}\int d\Omega_n^{(\hat k)}\sum_{p=1}^n
|\alpha(p,\hat k,\omega;\lambda)|^2=\frac{1}{n\Omega_n}
\left(\frac{2\pi}{\omega}\right)^n\,.
\ee
\subsection*{Gravitational perturbations}
In our normalizations, a spin-two field $\Phi_{\mu\nu}$ has number flux
$\Phi^*_{\mu\nu}\Phi^{\mu\nu}$. (This choice is possible because the waves are monocromatic, i.e.\
$\sim e^{{\rm i}\omega t}$.) If $\Phi_{\mu\nu}$ is multiplied by a suitable factor, it can be
interpreted as a metric perturbation $h_{\mu\nu}$. The multiplication factor depends on $\omega$
and $\hbar$, but its explicit form is not relevant in the derivation of the cross section.
An incoming transverse spin-two plane wave $\phi_{\mu\nu}^{(p,\hat k,\omega)}$
with unit flux and wave vector at infinity $\vec k=\omega\hat k$ is
\be
\phi_{\mu\nu}^{(p,\hat k,\omega)}
\rightarrow e^{-{\rm i}\omega t-{\rm i}\vec k\vec x}
\left(\begin{array}{c|c} 0 & 0 \\ \hline  0& \epsilon^p_{ij}(\hat k) \\ \end{array} \right)\,,
\ee
where $\epsilon^p_{ij}(\hat k)$ are the transverse traceless polarization vectors,
$p=1,\dots\,{\cal N}=(n-1)(n+2)/2$ and
\be
\hat k^i\epsilon^p_{ij}=0\,,\qquad
\epsilon^p_{ij}\epsilon^{p'}_{kl}\delta^{ik}\delta^{jl}=\delta^{pp'}\,,\qquad
\sum_{p=1}^{\cal N}\epsilon^p_{ij}\epsilon^{p\,kl}=
\delta_{(i}^k\delta_{j)}^l-\frac{1}{n}\delta_{ij}\delta^{kl}
+\hbox{ terms in }k_i,k_j,k^k,k^l\,.
\ee
This wave can be expanded in spherical waves $\Psi_{\mu\nu}^{(\lambda,\omega)}$
\be
\phi_{\mu\nu}^{(p,\hat k,\omega)}=\sum_\lambda\alpha(p,\hat k,\omega;\lambda)
\Psi_{\mu\nu}^{(\lambda,\omega)}\,.
\ee
The asymptotic fields $\Psi_{\mu\nu}$ are
\be
\Psi_{\mu\nu}^{(\lambda,\omega)}\rightarrow
\frac{e^{-{\rm i}\omega t-{\rm i}\omega r}}{r^{n/2}}
\left(\begin{array}{c|c} \begin{array}{cc} 0 &  \\  & 0 \\ \end{array}
 & 0 \\ \hline 0  &  r^2Y^{(\lambda)}_{ab} \\ \end{array}\right)\,,
\ee
where $Y_{ab}^{(\lambda)}$ satisfy the normalization and orthogonality conditions
\be
\int
d\Omega_nY_{ab}^{(\lambda)}Y_{cd}^{(\lambda')}\gamma^{ac}\gamma^{bd}
=\delta^{\lambda\lambda'}\,.
\ee
From the above equation it follows
\be
r^n\int d\Omega_n \Psi_{\mu\nu}^{*(\lambda,\omega)}\Psi^{(\lambda,\omega)~\mu\nu}=1\,,
\ee
i.e.\ a wave $\Psi_{\mu\nu}^{(\lambda,\omega)}$ describes one ingoing particle per unit time.  If
${\cal A}^{s=2}(\lambda,\omega)$ is the absorption coefficient for a spherical wave with angular
profile $Y_{ab}^{(\lambda)}$, the cross section associated to the plane wave $(p,\hat k,\omega)$ is
\be
\sigma^{p,\hat k,\omega}=\sum_\lambda|\alpha(p,\hat k,\omega;\lambda)|^2
|{\cal A}^{s=2}(\omega,\lambda)|^2\,.
\ee
The average on the wave direction and polarization of the initial state is obtained by summing on
$p$, integrating $\hat k$ on the sphere and dividing by the unit volume $\Omega_n$ and the number
of polarizations ${\cal N}$:
\be
\sigma=\frac{1}{{\cal N}\Omega_n}\sum_\lambda\int d\Omega_n^{(\hat k)}
\sum_{p=1}^{\cal N}|\alpha(p,\hat k,\omega;\lambda)|^2|{\cal A}^{(\omega,\lambda)}|^2\,.
\ee
From the previous equation, we obtain
\be
C^{s=2}(n+2,\omega)=\frac{1}{{\cal N}\Omega_n}\int d\Omega_n^{(\hat k)}
\sum_{p=1}^{\cal N}|\alpha(p,\hat k,\omega;\lambda)|^2\,.
\ee
The coefficients $\alpha$ are the coefficients of the expansion
\be
\tilde\phi_{\mu\nu}^{(p,\hat k,\omega)}=\sum_\lambda\alpha(p,\hat k,\omega;\lambda)
\tilde\Psi_{\mu\nu}^{(\lambda,\omega)}\,,
\ee
where
\be
\tilde\Psi_{ij}^{(\lambda,\omega)}\equiv
\frac{e^{-{\rm i}\omega t-{\rm i}\omega r}}{r^{n/2}}
\left(\begin{array}{c|c} 0
 & 0 \\ \hline 0  &  r^2Y^{(\lambda)}_{ab} \\ \end{array}\right)\,,\qquad
\tilde\phi_{ij}^{(p,\hat k,\omega)}\equiv e^{-{\rm i}\omega t-{\rm i}\vec k\vec x}
\epsilon^p_{ij}(\hat k)
\ee
are defined in the flat $(n+1)$-dimensional Euclidean space. From
\be
\int d^{n+1}x\tilde\Psi_{ij}^{*(\lambda',\omega')}\tilde\Psi^{ij(\lambda,\omega)}
=2\pi\delta(\omega-\omega')\delta_{\lambda\lambda'}\,,
\ee
it follows
\be
(2\pi)^2\delta(\omega-\omega')\delta(\omega'-
\omega'')|\alpha(p,\hat k,\omega;\lambda)|^2
=\int d^{n+1}xd^{n+1}x'\tilde\Psi_{ij}^{*(\lambda,\omega)}(x)
\tilde\phi^{ij(p,\hat k,\omega')}(x)\tilde\phi_{kl}^{*(p,\hat k,\omega')}(x')
\tilde\Psi^{kl(\lambda,\omega'')}(x')\,.
\ee
Multiplying by $\omega^{\prime n}$, summing over $p$ and integrating in $d\Omega_n^{(\hat
k)}d\omega'$, we obtain
\beq
&&\omega^n(2\pi)^2\delta(\omega-\omega'')\int d\Omega^{(\hat k)}_n
\sum_{p=1}^{\cal N}|\alpha(p,\hat k,\omega;\lambda)|^2\nonumber\\
&=&\int d^{n+1}xd^{n+1}x'\tilde\Psi_{ij}^{*(\lambda,\omega)}(x)
\tilde\Psi^{kl(\lambda,\omega'')}(x')\left[\int d\omega'\omega^{\prime n}
\int d\Omega^{(\hat k)}_n\left(\sum_{p=1}^{\cal N}\epsilon^{ij(p)}(\hat k)
\epsilon^{*(p)}_{kl}(\hat k)\right)
e^{-{\rm i}\vec k(\vec x-\vec x')}\right]\nonumber\\
&=&(2\pi)^{n+1}\int d^{n+1}xd^{n+1}x'\delta(\vec x-\vec x')
\tilde\Psi_{ij}^{*(\lambda,\omega)}(x)\tilde\Psi^{kl(\lambda,\omega'')}(x')
\left(\delta^{(i}_k\delta^{j)}_l-\frac{1}{n}\delta^{ij}\delta_{kl}
+\hbox{ terms in }k^i,k^j,k_k,k_l\right)\nonumber\\
&=&
(2\pi)^{n+2}\delta(\omega-\omega'')\,.
\nonumber\\
\eeq
Finally, the normalization factor is
\be C^{s=2}(n+2,\omega)=\frac{1}{{\cal N}\Omega_n}\int d\Omega_n^{(\hat k)} \sum_{p=1}^{\cal N} |\alpha(p,\hat
k,\omega;\lambda)|^2=\frac{1}{{\cal N}\Omega_n} \left(\frac{2\pi}{\omega}\right)^n\,. \ee
%

\end{document}